\documentstyle[fleqn,epsfig]{article}  

\def\be{\begin{equation}} 
\def\ee{\end{equation}} 
\def\bea{\begin{eqnarray}}
\def\eea{\end{eqnarray}}
\def\beqa{\begin{eqnarray*}}
\def\eeqa{\end{eqnarray*}}
\def\nn{\nonumber}
\begin{document}

\begin{titlepage}

\large
\centerline {\bf Bounds on Direct Couplings of Superheavy Metastable Particles 
to the }
\centerline{ \bf Inflaton Field from Ultra High Energy Cosmic Ray Events }
\normalsize
\vskip 1.0cm 

\centerline{A. H. Campos$^a$~\footnote{hcampos@charme.if.usp.br},
L. L. Lengruber$^b$~\footnote{leticia@ift.unesp.br}, 
 H. C. Reis$^c$~\footnote{hreis@ifi.unicamp.br}, R.
Rosenfeld$^b$~\footnote{rosenfel@ift.unesp.br}, 
and 
R. Sato$^c$~\footnote{rsato@ifi.unicamp.br}   }
\smallskip
\vskip 1.0cm

\centerline{$^a$ \it Instituto de F\'{\i}sica, DFN  - USP} 
\centerline{\it Rua do Mat\~ao , Trav. R, 187}
\centerline{\it 05508-900 S\~ao Paulo, SP, Brazil} 
\vskip 0.5cm 

\centerline {$^b$ \it Instituto de F\'{\i}sica Te\'orica - UNESP}
\centerline{\it Rua Pamplona, 145}
\centerline{\it 01405-900 S\~ao Paulo, SP, Brazil}
\vskip 0.5cm

\centerline{ $^c$ \it Departamento de Raios C\'osmicos e Cronologia}
\centerline{\it Universidade Estadual de Campinas - Unicamp}
\centerline{ \it CP 6165, 13083-970, Campinas - SP, Brazil}  

\vskip 2.0cm

\centerline {\bf Abstract}
\vskip 1.0cm
   
Top-down models for the origin of ultra high energy cosmic rays (UHECR's)
propose that these events are the decay products of relic superheavy metastable
particles, usually called $X$ particles. These particles can be produced in the
reheating period following the inflationary epoch of the early Universe.
We obtain constraints on some parameters such as the
lifetime and direct couplings of the $X$-particle to the inflaton field from the
requirement that they are responsible for the observed UHECR flux.
\bigskip

\noindent
PACS Categories:   98.70.Sa, 98.80.Cq

\vfill
\end{titlepage}

\newpage

\section{Introduction}

The ultra high energy cosmic rays (UHECR's), events with primary energies above 
 $E>10^{20}$\,eV, 
constitute one of the most intriguing mysteries in nature. Their origin, 
source and composition are not understood since they were first observed by the 
Volcano Ranch experiment in 1963 \cite{VR}. 

According to conventional theories, cosmic rays are produced by the acceleration
of charged particles in intense electromagnetic fields of astrophysical objects. This 
conventional wisdom, however, has some difficulties in explaining the observed flux of 
UHECR's. Most of the known astrophysical objects do not have an electromagnetic field strong 
enough neither a large acceleration site capable to accelerate particles up to
$10^{19}$ eV \cite{HP}--\cite{FE}.

Another problem is the propagation of the UHECR's through the intergalactic medium. If the
particles are
protons or nuclei, they should quickly degrade their energy due to interactions 
with the cosmic microwave background (CMB), producing a cut in the spectrum at
energies above $3 \times 10^{9}$ eV, the so-called 
GZK limit \cite{G}--\cite{F}. Simulations \cite{Cronin} show that in order that protons with 
energy larger than $10^{20}$\,eV arrive at Earth they must have traveled less than
$\approx\!\!100$\,Mpc, the attenuation length. A cut off is also expected if the primary 
particles are high energy photons in the range of $3\times 10^{12}$ eV to $10^{22}$ eV,
which have attenuation length smaller than $100$ Mpc, due to 
their interaction with various background photons (infrared/optical, CMB or universal radio 
background), mainly by pair production and inverse Compton scattering. These 
consecutive interactions produce electromagnetic cascades with decreasing energy that 
end up below $100$ GeV resulting in a gamma-ray background.

Furthermore, UHECR's are usually not deflected by intergalactic magnetic fields, so their 
incoming directions
should point to the sources.
However, taking into account the arrival directions of the UHECR 
events \cite{direction} and
constraining their origin to be in a distance of $100$ Mpc around the Earth, 
no candidate sources were found so far (an exception is a suggestion that these
events can be generated in M87 galaxy in the Virgo Cluster and made isotropic by
the galactic magnetic field \cite{tanco}).

A new scenario, the so-called top-down mechanism, was then proposed  
to explain these events \cite{Sigl}. 
In this scenario, UHECR's would originate from the decay of some superheavy 
particles ($m_X\!\!\gg\!\!10^{20}$\,eV) inside the GZK volume. They could be either emitted by
topological defects at the present epoch \cite{topological} or created in the pos-inflationary
era \cite{relics}. In the second case, where the supermassive particles must be
metastable ($\tau_X \geq 10^{10}$\,years), they could have been created perturbatively, by
a mechanism that is called {\it reheating} \cite{dw}, or by no-perturbative 
process like {\it preheating} \cite{preheating}, {\it instant preheating} 
\cite{FKL} or {\it gravitational production} \cite{Ford}. 
In this letter we will study bounds 
on the direct coupling of $X$-particles produced during the reheating phase
to the inflaton field by requiring that they are  sources of UHECR's. 
A preliminary version of this work appeared in \cite{campinas}. 

\section{Observed UHECR's flux and $X$-particle abundance}

We will now derive a relation among the abundance $\Omega_X$, the lifetime 
$\tau_X$ and the mass $m_X$ of the $X$-particle in order to provide the observed
flux of UHECR's on Earth. 

It is reasonable to assume that $X$-particles decay into
quarks and leptons. Particles are produced following the quark-hadron 
fragmentation process
described by QCD that results in the production of nucleons ($\approx 10\%$) and pions 
($\approx 90\%$)
distributed equally among the three charge states. Charged pions produce basically
electrons and neutrinos and neutral pions decay into photons. Photons dominate
the primary spectrum over protons by a factor of $\approx 6$ \cite{spectrum}
--\cite{spectrum1}. 
Following Sigl and 
Bhattacharjee \cite{Sigl}, we 
will consider that the UHECR's are photons with energy $E\simeq 10^{20}$\,eV and an 
attenuation length of $l(E_\gamma)\approx10$\,Mpc. It was recently suggested
that photons can indeed provide the bulk of UHECR's without violating EGRET bounds
on GeV gamma-ray backgrounds  \cite{photons}.

The photon flux generated by the decay of an uniform distribution of $X$-particles
is given by:
\be J_{\gamma} =
\frac{1}{4\pi}\,l(E_{\gamma})\,\dot{n}_X\,\frac{dN_\gamma}{dE_{\gamma}}\;, 
\label{fluxo}\ee
where $dN_\gamma/dE_\gamma$ is the photon injection spectrum from $X$-decay and 
$\dot{n}_X (=n_X/\tau_X)$ is the $X$-decay rate.

From the observed UHECR's flux \cite{Sigl}, 
$J(E\simeq 10^{20}\,\mbox{eV}) = 10^{-27}\,(\mbox{m}^2\,\mbox{sr\,s\,GeV)}^{-1}$
and using $dN_\gamma/dE_\gamma \approx E_{\gamma}^{-1.5}$ from a QCD model \cite{Sigl}, one 
gets:
\be
\dot{n}_X = 3.7\times10^{-46}\left(\frac{m_X}
{10^{12}\,\mbox{GeV}}\right)^{-1/2}\mbox{cm}^{-3}\mbox{s}^{-1}\;.
\label{ndot}\ee
So, we can obtain a relation between $\Omega_Xh^2$ and $\tau_X$: 
\be\tau_X\!\!=\!\frac{\Omega_X\rho_c}{\dot{n}_X m_X}\!\simeq\!10^{22}(\Omega_X
h^2)\!\left(\frac{10^{12}\mbox{GeV}}{m_X}\right)^{1/2}\!\!\!\mbox{years,}
\label{Omega}\ee
where $\Omega_X = \rho_X/\rho_c$, $\rho_X$ is the $X$-particle mass density, 
$\rho_c \simeq 10^{-5}h^{2}$\,GeV\,cm$^{-3}$  is the critical density,
and $h$ is the present value of Hubble's constant in units of
$100$\,km\,s$^{-1}$Mpc$^{-1}$.

It could be interesting that these unknown $X$-particles constitute 
a fraction of the cold dark matter 
(CDM). In this case, they
could be concentrated in galactic haloes and should not be uniformly 
distributed along the whole 
universe, as assumed in (\ref{fluxo}).
Considering just the contribution of our galactic halo the correct equation 
to describe the flux 
of UHECR's is:
\be J^{h}_{\gamma} =
\frac{1}{4\pi}\,R^{h}\,\dot{n}_{X}^{h}\,\frac{dN_\gamma}{dE_{\gamma}}\;,\ee
where $R^{h}$ is the radius of the galactic halo and
$\dot{n}_{X}^{h}$ is the decay rate of the $X$-particles clustered in the halo.

The flux above can be written in terms of $J_{\gamma}$, given in (\ref{fluxo}),
so that $J^{h}_{\gamma}=fJ_{\gamma}$ and:  

\bea
f&=& \frac{J^{h}_{\gamma}}{J_{\gamma}} = \frac{n_{X}^{h}R^{h}}{n_X
l(E_{\gamma})} = \nn  \\
&=& 1.5 \times 10^3
\left(\frac{0.2}{\Omega_{CDM}h^2}\right) \left(\frac{R^{h}}{100\,
\mbox{Kpc}}\right)\left(\frac{10\,\mbox{Mpc}}{l(E_\gamma)}\right) 
\left( \frac{\rho^h_{CDM}}{0.3\:\mbox{GeV\,cm}^{-3}}\!\!\right)\nn
\eea 
where $\rho^h_{CDM}$ is the cold dark matter energy density in the halo. 
We recall that if 
$X$-particles are a fraction of CDM, it can be written as 
$\rho^h_X=\varepsilon \rho^h_{CDM}$.  The fraction $\varepsilon$ is the 
same everywhere in the 
universe, so that $\varepsilon=\Omega_X/\Omega_{CDM}$.

The halo concentration of $X$-particles implies a modification of equation 
(\ref{Omega}):
\be
\tau_X = 10^{22}\,f\,\Omega_X
h^2\left(\frac{10^{12}\,\mbox{GeV}}{m_X}\right)^{1/2}\;\mbox{years ,}
\label{tau}
\ee
\vskip 0.3cm
Considering $m_X=10^{12}$\,GeV, for a minimum  $\tau_X$ of
$10^{10}$\,years and a maximum $\Omega_Xh^2$ of the order of $1$, we obtain
the following limits (for $f \approx 10^{3})$ (see also \cite{spectrum1}):

\begin{equation} 
\begin{array}{lll}
{\mbox{for}}\,\,\, (\Omega_X h^2) \simeq 1   &\,\rightarrow& \, \tau_X  \simeq
10^{25}\,\mbox{years,} \nonumber \\
{\mbox{for}}\,\,\, \tau_X \simeq 10^{10}\,\mbox{years} &\, \rightarrow& \,(\Omega_X h^2)  \simeq
10^{-15} \;\; . 
\label{limits}
\end{array} 
\end{equation}
Therefore, cosmologically interesting abundances of X-particles which at the
same time provide a solution to the UHECR events require a very large 
life-time $\tau_X$. These requirements are generally difficult to obtain in
realistic theories, but examples do exist \cite{examples}.

\vskip 1.0cm
\section{Production of $X$-particles in the reheating phase} 

In the inflationary scenario, following the exponential expansion of the 
universe, particles  are created by the
oscillation of inflaton field $\phi$ about its minimum. The coupling of the 
inflaton with other 
fields becomes important
and allows the transference of its energy to other particles. 
In the perturbative reheating process  \cite{dw},
the inflaton decays mostly into relativistic particles (radiation) 
that reheat the  universe.  
In this context, the superheavy particle $X$ could be 
created by either radiation annihilation process (indirect production) 
or by direct
inflaton decay $\phi\rightarrow X\bar{X}$ (direct production), if such a 
coupling is allowed.
The study of the latter mechanism is the main focus of this work.  

In order to study quantitatively $X$-particle production during reheating, 
we have
to solve the coupled differential Boltzmann equations for the energy densities
of $X$ particles $(\rho_X)$, radiation  $(\rho_R)$ and inflaton $(\rho_{\phi})$
:

\be
\dot{\rho_\phi} + 3H\rho_\phi + \Gamma_\phi = 0,
\ee
\be
\dot{\rho_R} + 4H\rho_R - (1-B_X)\Gamma_{\phi} \rho_{\phi} -  
\frac{\langle\sigma|v|\rangle}{m_X}[(\rho_X)^2 - (\rho^{eq}_X)^2] = 0,  
\label{system}
\ee  
\be
\dot{\rho_X} + 3H\rho_X - \frac{m_X}{m_{\phi}}B_X\Gamma_{\phi} 
\rho_{\phi} + \frac{\langle\sigma|v|\rangle}{m_X}[(\rho_X)^2 
- (\rho^{eq}_X)^2] = 0,
\label{X}
\ee
where:
\be
H^2=\frac{8\pi}{m_{pl}^2}(\rho_{\phi}+ \rho_R +\rho_X).
\label{H}
\ee
In the above equations $H$ is the expansion rate of the universe, $B_X$ is 
the branching ratio for
the $\phi \rightarrow X \bar{X}$ decay, $\rho^{eq}_X$
is the equilibrium value for the $X$-particle energy density and $\langle\sigma|v|\rangle$ is the 
thermal average of
the $X$ annihilation cross section times the M{\o}ller flux factor.
The third term in equation (\ref{X}) has a factor $m_X/m_{\phi}$ that
corrects a similar equation in \cite{Chung}. 
According to \cite{kolb}, 
the annihilation cross section can be estimated as   
$\langle\sigma|v|\rangle\sim \alpha^2/T^2$ where $g=\sqrt{4\pi\alpha}$ is the gauge coupling 
strength and $T$ is the characteristic temperature for the process. 
$\Gamma_\phi$
is the inflaton total width 
and is
obtained as a function of the reheating temperature, $T_{RH}$, assuming
instantaneous conversion of the inflaton energy density into radiation:

\be
\Gamma_{\phi} = 1.93 \times 10^{-18} \left(\frac{T_{RH}}{\mbox{GeV}}\right)^2
\,\mbox{GeV}\;.
\ee
It is important to notice that the reheating temperature ($T_{RH}$) 
is not the highest 
temperature achieved by the universe during reheating. For example, 
in this paper we are using a 
simple chaotic inflation model where $V= m_{\phi}^2 {\phi}^2/2$ and 
$m_{\phi}=10^{13}$\,GeV. 
Following \cite{Chung} it can be shown  that, even if $T_{RH} = 10^{9}$ GeV, 
the universe can achieve 
temperatures as high as $10^{12}$ GeV.   In this case, for $g=1$ we have  
$\langle\sigma|v|\rangle = 10^{-26}$\,GeV$^{-2}$, which allows the process  
$\gamma \gamma\rightarrow X\bar{X}$.

It is convenient to work with co-moving and dimensionless variables and
therefore we defined appropriate parameters \cite{Chung} and solved the 
equations numerically for the scaled variable
$X\equiv\rho_{X}{m_X}^{-1}a^3$, where $a(t)$ is the scale factor. In this
case, the new independent variable is $x=am_{\phi}$.

In Figure 1, we illustrate the behavior of $X(x)$.  
The curves A, B and C  show two plateaus 
associated with two mechanisms of X-particle
production: the first one is the indirect production 
($\gamma\gamma \rightarrow X\bar{X}$) and 
the second is the direct one ($\phi \rightarrow X\bar{X}$).
Modifying $B_X$ and keeping $T_{RH}$ constant (B and C curves) 
only the second plateau changes, 
reflecting the
dependence of the direct decay on $B_X$. On the other hand, 
when $T_{RH}$ is much smaller than  
$10^{9}$ GeV, there is not enough energy for the indirect 
process and hence, the first plateau 
tends to disappear.

\begin{figure}[t]
\epsfig{file=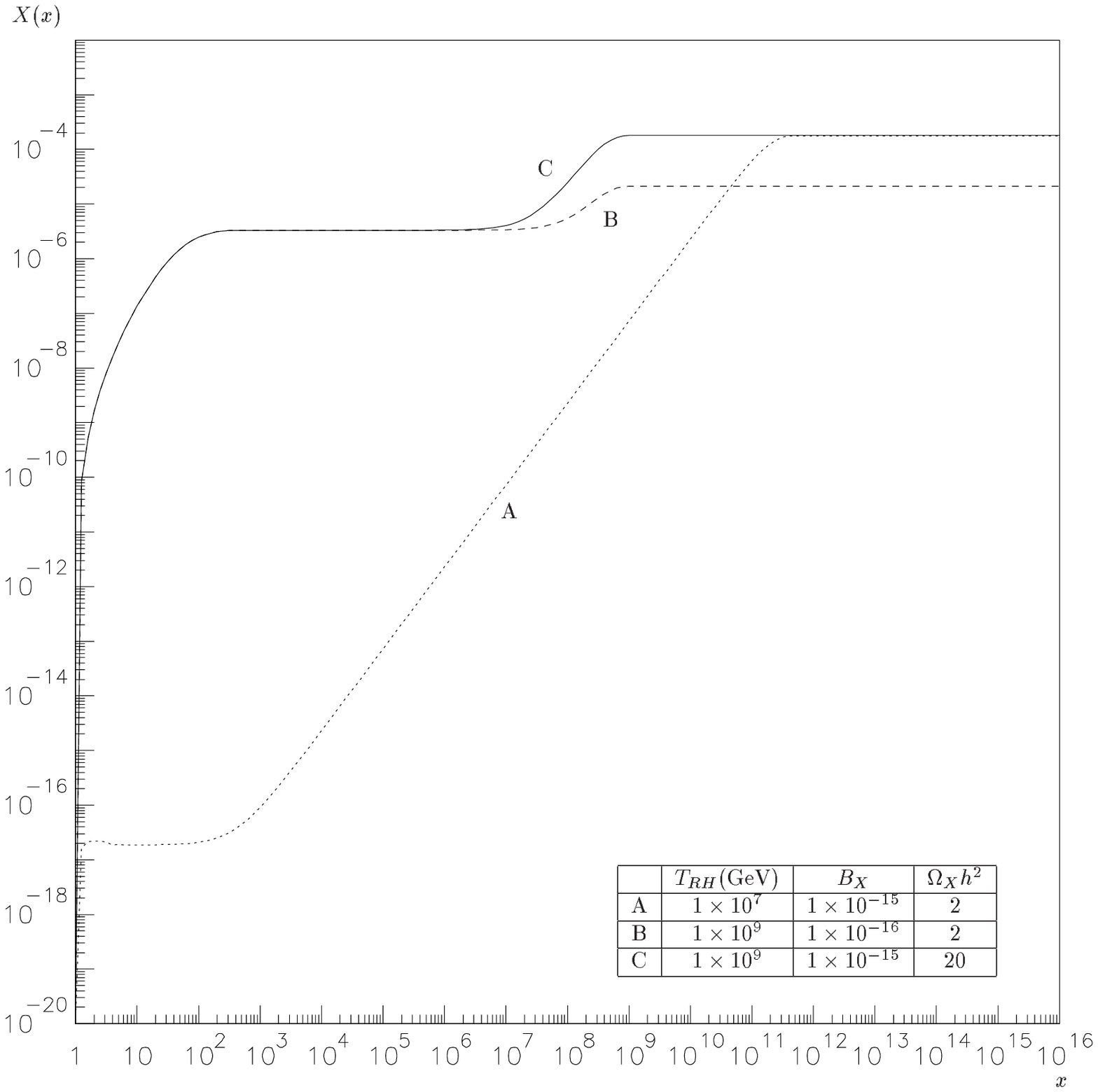,
        bbllx= 2.47cm,
        bblly= 7.38cm,
        bburx=19.20cm,
        bbury=24.78cm,
        width=\linewidth
}
\caption{Evolution of $X$ ($\equiv\rho_{X}
m_X^{-1}a^3$) as a function of the modified scale factor, $x$.
We used $m_{\phi}=10^{13}$\,GeV
and $m_X=10^{12}$\,GeV.}
\label{graf1}
\end{figure}
   
\section{Finding $\Omega_X$}   

We can find the $X$-particle mass density $\Omega_X$ today using
the present value ($x = x_0$) of the
radiation abundance $\Omega_R$ $(\Omega_Rh^2 = 4.3 \times 10^{-5})$:
\be
\frac{\Omega_Xh^2}{\Omega_Rh^2} = \frac{\rho_X(x_0)}{\rho_R(x_0)}
\label{Omega1}
\ee

In thermal equilibrium, $T\propto {g_*}^{-1/3}a^{-1}$ \cite{kolb}, and hence
$\rho_R$ can be written as:
\be
\rho_{R}(x_0)=\rho_{R}(x_{RH})\left(\frac{g_{*}(x_{RH})}{g_{*}(x)}\right)^{1/3}
\left(\frac{x_{RH}}{x_0}\right)^4 
\ee
where $g_*$ is the effective number of degrees of freedom (we used $g_*(x_{RH})
= 200$).

Assuming that $X$ particles are created already decoupled, we can write that 
$\rho_X(x_0)=\rho_X(x_{RH})(x_{RH}/x_0)^{3}$, where:
\be  \rho_{X}(x_{RH}) = X_{F}
{x_{RH}}^{-3} m_{\phi}m_X. 
\label{rhoX_Xf}
\ee

The parameter $X_F$ is the $X$ value in the second plateau (Figure 1) and 
corresponds  to $X(x_{RH})$.
So, we get:
\bea
\Omega_{X}h^2 &=& \Omega_{R}h^2 \frac{\rho_{X}(x_{RH
})}{\rho_{R}(x_{RH})}
\frac{T_{RH}}{T_0} \nn \\
&=& 1.5 \times 10^{9} \left(\frac{T_{RH}}{\mbox{GeV}}\right) \frac
{X_{F}m_{X}m^{3}_{\phi}}{x^{3}_i H^{2}_{i}m^{2}_{pl}},
\label{Omega2}
\eea
where $m_{pl}$ is the Planck mass and $T_0=2.37\times 10^{-13}$\,GeV is the
present-epoch CMB temperature. 

\section{Numerical Results}

We worked with acceptable reheating temperatures in the range of 
$10^{-2} \mbox{GeV} \leq T_{RH} \geq 10^9 \mbox{GeV}$. The lower limit is imposed 
by nucleosyntesis
and the upper one is chosen not to lead to overproduction of gravitinos 
\cite{gravitino}.  
The subscript $i$ refers to the epoch before inflaton decay. For
chaotic inflation models we have $H_{i} \approx m_{\phi}$. As initial 
condition for
the differential equations we choose $x_i=1$.

Solving the Boltzmann equations numerically for different $B_X$ and $T_{RH}$, 
we obtain the 
respective $X_F$ and consequently its associated $\Omega_Xh^2$, 
as it is shown in the table 1.

\begin{table*}
\begin{tabular}{|c||c|c||c|c||c|c|}
\hline
&
\multicolumn{2}{c||}{$T_{RH}=10^9\,\mbox{GeV}$}
&\multicolumn{2}{c||}{$T_{RH}=10^3\,\mbox{GeV}$}
&\multicolumn{2}{c|}{$T_{RH}=10^{-2}\,\mbox{GeV}$}\\ 
\cline{2-7}
\raisebox{1.5ex}[0pt]{$B_{X}$} &    $X_{F}$ & $\Omega_{X} h^{2}$ &    $X_{F}$ & $\Omega_{X} h^{2}$ &    $X_{F}$ & $\Omega_{X} h^{2}$   \\
\hline
$  0  $ & $ 3.3\times 10^{-6}$          &   $  0.38  $       &$  0  $          &   $  0  $       &$  0  $          &   $  0  $       \\

\hline
$ 1\times 10^{-15}$&$ 1.8\times 10^{-4}$          &   $20.97$       &$ 1.78\times 10^{-4}$      
   &   $2.06\times 10^{-5}$       &$ 1.78\times 10^{-4}$          &   $2.06\times 10^{-10}$       \\

\hline
$ 1\times 10^{-11}$&$ 1.78 $          &   $2.06\times 10^{5}$       &$ 1.78$          &  
$2.06\times 10^{-1}$       &$ 1.78 $          &   $2.06\times 10^{-6}$       \\

\hline
$1\times 10^{-9}$&$ 1.78\times 10^{2}$          &   $2.06\times 10^{7}$       &$ 1.78\times
10^{2}$          &   $2.06\times 10^{1}$       &$ 1.78\times 10^{2}$          &   $2.06\times 10^{-4}$       \\

\hline
$1\times 10^{-5}$&$ 1.78\times 10^{6}$          &   $2.06\times 10^{11}$       &$ 1.78\times
10^{6}$          &   $2.06\times 10^{5}$       &$ 1.78\times 10^{6}$          &   $ 2.06 $       \\

\hline
$1\times 10^{-3}$&$ 1.78\times 10^{8}$          &   $2.06\times 10^{13}$       &$ 1.78\times
10^{8}$          &   $2.06\times 10^{7}$       &$ 1.78\times 10^{8}$          &   $2.06\times 10^{2}$       \\
\hline 
\end{tabular}
\caption{Numerical results for the final abundance obtained by solving the 
Boltzmann equations, for $T_{RH} = 10^9, 10^3, 10^{-2}$ GeV and  $B_X < 10^{-3}$.}
\end{table*}

For $B_X=0$, the $X$ particle production is not effective if the 
reheating temperature is lower than $T=10^9$ GeV.  From table 1 we also 
get the expressions 
below, related to the direct
production process that prevails over the indirect one, for $B_X > 10^{-16}$:
\bea
X_F &=& 1.78 \times 10^{11} B_X,  \\ 
\Omega_Xh^2 &\simeq& 2.06 \times 10^7 \left(\frac{T_{RH}}{\mbox{GeV}}\right) B_X.
\label{Omega4}
\eea 

An excellent analytical approximation to the numerical results can be obtained 
by requiring that $n_X = n_{\phi} B_X$, in which case we find: 
\be
X_F = \frac{3}{8\pi} \frac{m_{pl}^2}{m_{\Phi}^2} B_X
\ee
Substituting the numerical values the above equation reproduces 
(\ref{Omega4}) exactly. 

The maximum efficiency of the direct decay process 
($\phi \rightarrow X\bar{X}$) is obtained when 
$T_{RH}$ is minimum. Recently, low temperature reheating models were
explored in \cite{Giudice}. Taking for the minimum value of 
$T_{RH} \simeq 10^{-2}$ GeV and using the 
upper  limit for $X$ abundance ($\Omega_Xh^2 \simeq 1$) we get the maximum
value for the branching ratio of $B_X=10^{-5}$. 
 
More generally, we can find bounds on $B_X$ by requiring that 
$X$-particles are sources of UHECR's, in which case the limits of 
(\ref{limits}) apply. These bounds are shown in figure 2 as a function of the
reheating temperature for two limits on the abundance: 
$\Omega_Xh^2 \leq 1$  ($\tau_X \simeq 10^{25}$ years) and  
$\Omega_Xh^2 \geq 10^{-15}$ ($\tau_X \simeq 10^{10}$ years).

\begin{figure}[]
\begin{center}
\psfig{file=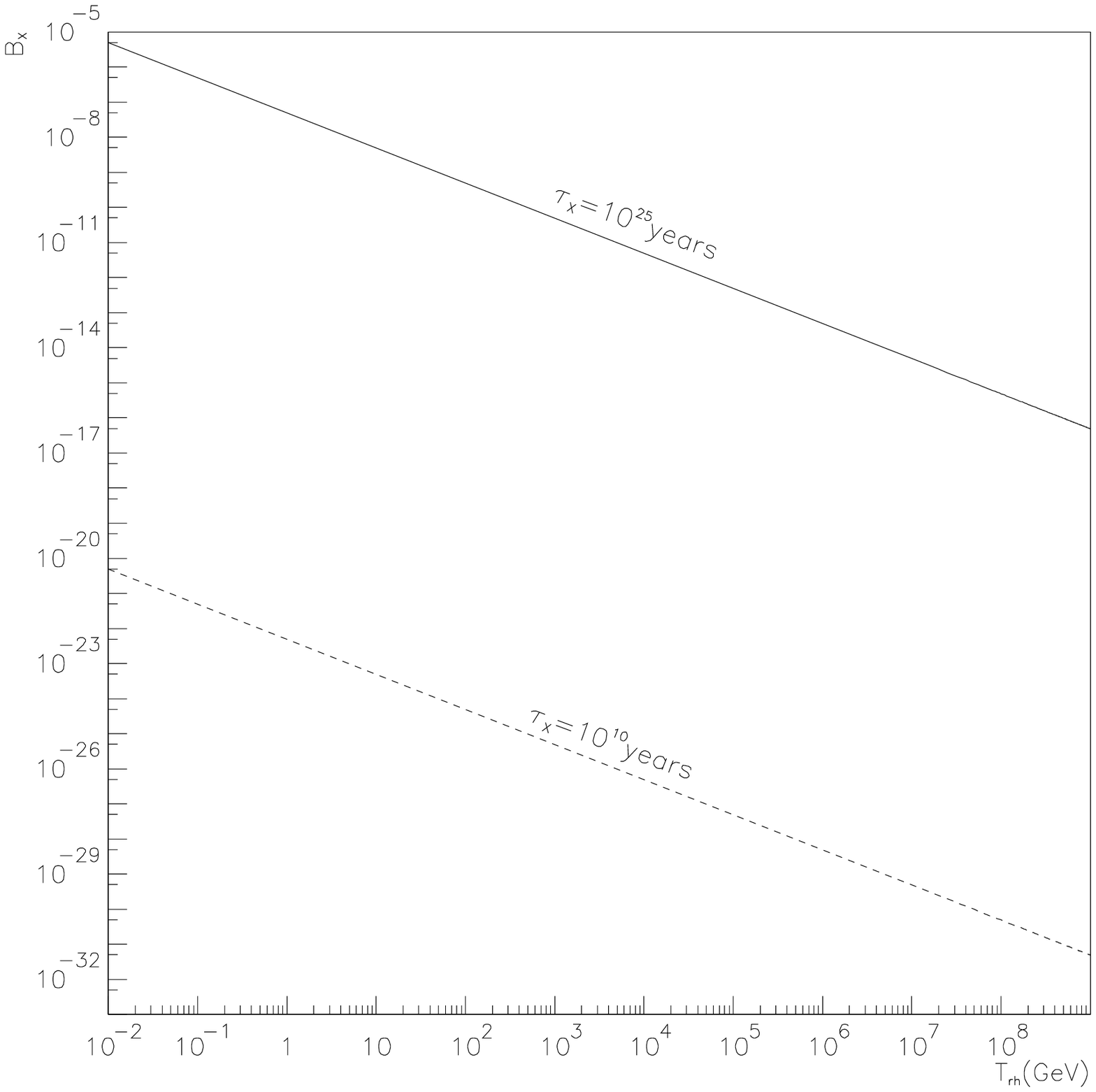,
        bbllx= 1.45cm,
        bblly= 5.51cm,
        bburx=18.60cm,
        bbury=23.05cm,
        width=\linewidth
}
\end{center}
\caption{Branching ratio $B_X$ of the inflaton into $X$-particles as a
function of
the reheating temperature $T_{RH}$ for two limiting values of $\Omega_X
h^2$.}
\label{graf2}
\end{figure}

\section{Conclusion}
 
We have investigated the constraints on the properties of supermassive relic
metastable particles demanding that their decay products are responsible for
the observed UHECR flux. In particular, we found the relevant range for the
branching ratio of inflaton decay into the $X$-particle as a function of the
reheating temperature.

For very small branching ratio of the inflaton to the $X$-particle, 
$B_X \geq 10^{-16}$, the direct production ($\phi\rightarrow X\bar{X}$) 
dominates over the indirect thermal production even for $T_{RH} = 10^{9}$ GeV.
For the smallest possible reheat temperature, $T_{RH} = 10^{-2}$ GeV, thermal
production of $X$-particle does not occur and the direct mechanism with 
$B_X \leq 10^{-5}$ is required.

Further consequences of the top-down scenario are the dominance of photons and
neutrinos as primaries of UHECR's \cite{spectrum} and an anisotropy of 
these events as a result of the large concentration of $X$-particles in the 
galactic center\cite{anisotropy}. 
More recently, it was suggested that clumps of dark matter in
the halo could explain the observed multiplets (12 doublets and 2 triplets) 
of UHECR's events \cite{clumps}.
A detailed analysis including several models
for galactic haloes of our and other galaxies was performed in \cite{halo}.
We look forward to new experimental results from current and future experiments
to elucidate the composition and anisotropy of the UHECR events. In particular,
the first site of the Auger Observatory in the southern hemisphere, being 
sensitive to the galactic center, will be important in testing the anisotropy 
resulting from the top-down scenario.  


\section*{Acknowledgments}

This work was supported by Funda\c{c}\~{a}o de Amparo \`{a} Pesquisa do 
Estado de S\~{a}o Paulo~(FAPESP) and Conselho Nacional de Desenvolvimento 
Cient\'{\i}fico e Tecnol\'{o}gico~(CNPq).

\end{document}